\documentclass[twocolumn,superscriptaddress]{revtex4-2}
\usepackage[]{graphicx}
\usepackage{textcomp}
\usepackage{amsmath}

\usepackage{mathtools}
\usepackage{physics}
\usepackage{siunitx}
\usepackage{braket}
\usepackage{verbatim}

\begin{document}

\title{Control of $\text{N}_{\text{2}}^{\text{+}}$ Air Lasing}

\author{Mathew Britton}
\email[]{mathew.britton@uottawa.ca}
\affiliation{Department of Physics, University of Ottawa, Ottawa, K1N 6N5, Canada}

\author{Marianna Lytova}
\affiliation{Department of Physics, University of Ottawa, Ottawa, K1N 6N5, Canada}

\author{Dong Hyuk Ko}
\affiliation{Department of Physics, University of Ottawa, Ottawa, K1N 6N5, Canada}

\author{Abdulaziz Alqasem}
\affiliation{Department of Physics, University of Ottawa, Ottawa, K1N 6N5, Canada}
\affiliation{Department of Physics, King Saud University, Riyadh, 11451, Saudi Arabia}

\author{Peng Peng}
\affiliation{Department of Physics, University of Ottawa, Ottawa, K1N 6N5, Canada}

\author{D. M. Villeneuve}
\affiliation{Department of Physics, University of Ottawa, Ottawa, K1N 6N5, Canada}
\affiliation{National Research Council of Canada, Ottawa, K1A 0R6, Canada}

\author{Chunmei Zhang}
\email[]{chunmei.zhang@uottawa.ca}
\affiliation{Department of Physics, University of Ottawa, Ottawa, K1N 6N5, Canada}

\author{Ladan Arissian}
\affiliation{Department of Physics, University of Ottawa, Ottawa, K1N 6N5, Canada}
\affiliation{National Research Council of Canada, Ottawa, K1A 0R6, Canada}
\affiliation{Center for High Technology Materials, Albuquerque, NM, 87106, USA}

\author{P. B. Corkum}
\email[]{pcorkum@uottawa.ca}
\affiliation{Department of Physics, University of Ottawa, Ottawa, K1N 6N5, Canada}

\date{\today}

\begin{abstract}
A near-infrared laser generates gain on transitions between the $\text{B}^{\text{2}} \Sigma_{\text{u}}^{\text{+}}$ and $\text{X}^{\text{2}} \Sigma_{\text{g}}^{\text{+}}$ states of the nitrogen molecular cation in part by coupling the $\text{X}^{\text{2}} \Sigma_{\text{g}}^{\text{+}}$ and $\text{A}^{\text{2}} \Pi_{\text{u}}$ states in the \textit{V}-system. Traditional time resolved pump-probe measurements rely on post-ionization coupling by the pump pulse to initialize dynamics in the $\text{A}^{\text{2}} \Pi_{\text{u}}$ state. Here we show that a weak second excitation pulse reduces ambiguity because it acts only on the ion independent of ionization. The additional control pulse can increase gain by moving population to the $\text{A}^{\text{2}} \Pi_{\text{u}}$ state, which modifies the lasing emission in two distinct ways. The presence of fast decoherence on $\text{X}^{\text{2}} \Sigma_{\text{g}}^{\text{+}}$ to $\text{A}^{\text{2}} \Pi_{\text{u}}$ transitions may prevent the formation of a coherent rotational wave packet in the ground state in our experiment, but the control pulse can reverse impulsive alignment by the pump pulse to remove rotational wave packets in the $\text{B}^{\text{2}} \Sigma_{\text{u}}^{\text{+}}$ state.
\end{abstract}

\pacs{42.65.Re}


\maketitle

\begin{figure*}[] 
\includegraphics[trim={0cm 0cm 0cm 0cm},clip]{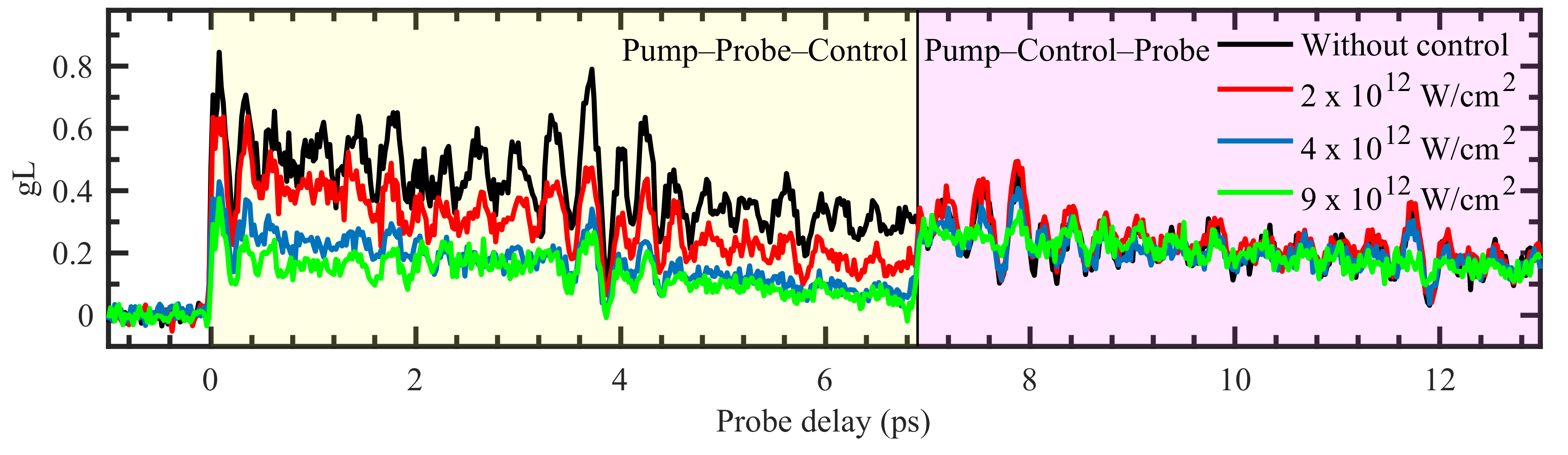}
\caption{\label{fig:intro} The gain-length product ($gL$) as a function of pump-probe delay. The black line is without a control pulse. When the control pulse arrives near 7~ps (vertical line), the experiment depends on the ordering of the pulses. The control pulse intensity ($I_{control}$) is in the legend. The pump intensity ($I_{pump}$) is $1.9\times10^{14}~\mathrm{W}/\mathrm{cm}^2$.}
\end{figure*}

\section{Introduction}

Focusing a powerful near-infrared femtosecond laser into air reveals gain on transitions in the ultraviolet between the different vibrational states of the upper $\text{B}^{\text{2}} \Sigma_{\text{u}}^{\text{+}}$ and the ground $\text{X}^{\text{2}} \Sigma_{\text{g}}^{\text{+}}$ electronic states of the nitrogen molecular cation~\cite{yao11}. Pump laser pulses near 800~nm move population from the ground state to the middle $\text{A}^{\text{2}} \Pi_{\text{u}}$ state of the ion~\cite{xu15,yao16,zhong17,richter17,xu18,li19,zhang19b,zhang19,ando19,wan19,zhang20}, which initiates a vibrational wave packet that can temporarily trap population~\cite{xu15,yao16,richter17}. The $\text{X}^{\text{2}} \Sigma_{\text{g}}^{\text{+}}$ to $\text{A}^{\text{2}} \Pi_{\text{u}}$ interaction contributes to gain by depleting the ground state, but it also enables control of the gain and emission. 

Elliptical~\cite{wan19} and polarization-modulated~\cite{li19,xie19,fu20} pump pulses exploit the post-ionization interaction with the $\text{A}^{\text{2}} \Pi_{\text{u}}$ state to control gain, but these experiments are complicated because the pump pulse acts before, during, and after ionization. A second weak excitation pulse can reduce ambiguity because it acts only on the ion independent of ionization. Electronic and vibrational coherences produce rapid oscillations of gain with multiple probe pulses~\cite{zhang19a} or pump pulses~\cite{ando19,mysyrowicz19,zhang19,chen19}. In addition, a second 800~nm pulse modifies gain and quenches emission, which are both consistent with the $\text{A}^{\text{2}} \Pi_{\text{u}}$ state interaction~\cite{zhang19}.

We perform a pump-probe experiment on the ion itself to investigate and harness the $\text{A}^{\text{2}} \Pi_{\text{u}}$ state. For clarity, we refer to the additional pulse as ``control.'' This experiment minimizes the interaction length using a narrow gas jet in vacuum to isolate gain from the effects of propagation~\cite{britton18,britton19}. By varying both control and probe delays, we also observe modified gain and emission. While the influence of the $\text{A}^{\text{2}} \Pi_{\text{u}}$ state on gain is clear, we discuss two distinct ways that its presence modifies the emission. The control pulse also strengthens or cancels the initial rotational wave packet in the $\text{B}^{\text{2}} \Sigma_{\text{u}}^{\text{+}}$ state with a new conventionally-generated rotational wave packet. The weakness of $\text{X}^{\text{2}} \Sigma_{\text{g}}^{\text{+}}$ state rotational frequencies allows us to conclude that the $\text{A}^{\text{2}} \Pi_{\text{u}}$ state interaction may prevent a coherent wave packet from forming in the $\text{X}^{\text{2}} \Sigma_{\text{g}}^{\text{+}}$ state.

\section{Results}

Figure~\ref{fig:intro} (black line) shows the gain-length product $gL$ as a function of probe delay for the traditional pump--probe measurement without a control pulse. Zero delay corresponds to the arrival of the pump pulse ({\raise.17ex\hbox{$\scriptstyle\mathtt{\sim}$}}35~fs, 800~nm), and the weak second harmonic probe pulse ({\raise.17ex\hbox{$\scriptstyle\mathtt{\sim}$}}100~fs, 400~nm, $<$$10^{10}$~W/cm$^2$) measures gain at 391~nm [$\text{B}^{\text{2}} \Sigma_{\text{u}}^{\text{+}}$ $\left(\nu = 0\right)$ $\rightarrow$ $\text{X}^{\text{2}} \Sigma_{\text{g}}^{\text{+}}$ $\left(\nu = 0\right)$]. The rotational wave packets in the states of the ion are imprinted on the decay of gain~\cite{kartashov14,richter20,zhang13,zeng14,xie14,lei17,zhong17,zhong18,arissian18,britton19}. 

When the weak control pulse ({\raise.17ex\hbox{$\scriptstyle\mathtt{\sim}$}}100~fs, 800~nm) arrives at a fixed delay in Fig.~\ref{fig:intro}, the experiment depends on the order of the three pulses: Pump--Probe--Control and Pump--Control--Probe. The latter is like the traditional experiment because the probe pulse measures dynamics initiated by both the pump and control pulses. As the control intensity increases in Fig.~\ref{fig:intro}, the amplitude of modulations from rotational wave packets decreases in Pump--Control--Probe measurements. In the Pump--Probe--Control measurements that we discuss later, the control pulse modifies the system during the emission following the probe pulse that is also known as free induction decay.

Figure~\ref{fig:wp}(a) shows the Fourier transform of the modulations without a control pulse (black line). The narrow peaks in the frequency domain identify wave packets from different electronic states. Vertical dotted lines mark the rotational frequencies for the upper $\text{B}^{\text{2}} \Sigma_{\text{u}}^{\text{+}}$ state, and they align well with the positions of the peaks. Conversely, few peaks align with the expected positions for the ground state of the ion and the neutral molecule, as shown in Supplementary Fig.~S1~\cite{SM}. The wave packet in the upper state is dominant in our experiment, which is consistent with fast decoherence in the $\text{X}^{\text{2}} \Sigma_{\text{g}}^{\text{+}}$ state or a large population inversion. 

\subsection{Pump--Control--Probe}

\begin{figure}[h!]
\includegraphics[trim={0cm 0cm 0cm 0cm},clip]{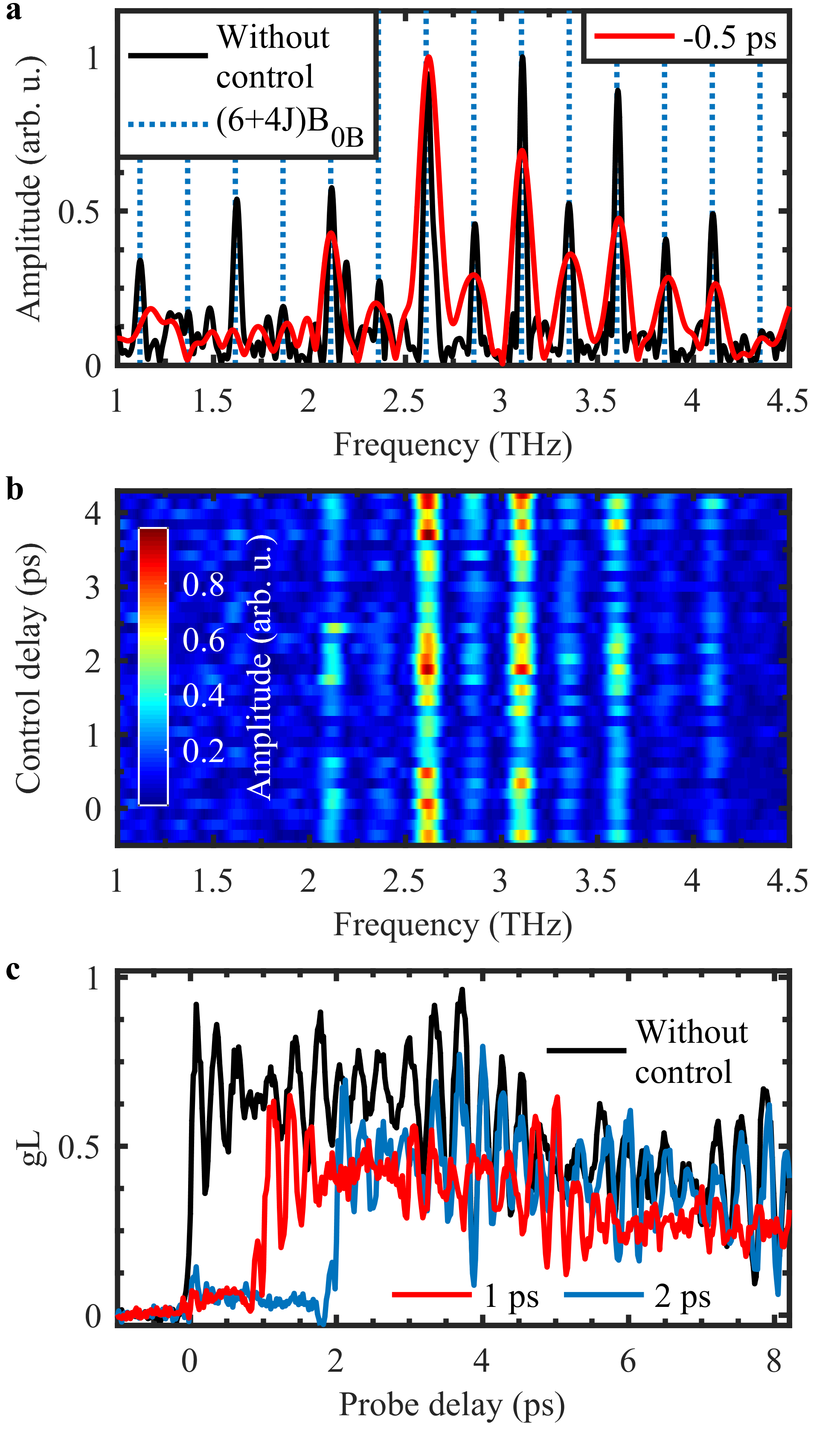}
\caption{\label{fig:wp} Rotational wave packets. (a) The amplitude of the Fourier transform of the modulations, where frequency is the Fourier conjugate parameter of the delay between the pump and probe. The black line is without the control pulse. The vertical dashed lines indicate the rotational frequencies for the $\text{B}^{\text{2}} \Sigma_{\text{u}}^{\text{+}}$ state. The lower-resolution data includes the control pulse at  $-0.5$~ps. (b) The amplitude of the modulations as a function of frequency and control delay. The peaks from the above panel oscillate as a function of control delay. $I_{control}$ {\raise.17ex\hbox{$\scriptstyle\mathtt{\sim}$}} $10^{13}~\mathrm{W}/\mathrm{cm}^2$. (c) An intense control pulse (${>}3\times10^{13}$~W/cm$^2$) at 1~ps and 2~ps. $I_{pump} = 2.5\times10^{14}~\mathrm{W}/\mathrm{cm}^2$.}
\end{figure}

The Fourier transform of the modulations shows how the rotational wave packets change. The red line in Fig.~\ref{fig:wp}(a) is the Fourier transform of modulations when the control pulse arrives before ionization ($-0.5$~ps), which is like the traditional measurement without a control pulse except for alignment of $\text{N}_{\text{2}}$ before ionization~\cite{xu17}. As the control delay increases beyond zero in Fig.~\ref{fig:wp}(b), an appropriately timed control pulse interaction periodically reduces or increases the amplitude of the $\text{B}^{\text{2}} \Sigma_{\text{u}}^{\text{+}}$ state rotational frequencies. To do this, the control pulse can create a new wave packet that interferes with the original~\cite{lee06}. This interference depends on the delay between the pump and control pulses.

If the control pulse intensity is high, Fig.~\ref{fig:wp}(c) shows that the new rotational wave packet can overpower the original and add new modulations that are offset in time. At 1~ps, the amplitude of the original modulations is reduced and modulations from the control pulse wave packet emerge. This confirms the role of rotational excitation by the control pulse and demonstrates a method to modify or minimize rotational wave packets in the $\text{N}_{\text{2}}^{\text{+}}$ air laser. The amplitude of the exponential decay beneath the modulations is lower with the control pulse in Fig.~\ref{fig:wp}(c), but we only observe this at high control intensity. At control intensities below about $3\times10^{13}$~W/cm$^2$, we usually observe higher gain (\textit{e.g.} Supplementary Fig.~S2~\cite{SM}). This is consistent with population exchange from the $\text{X}^{\text{2}} \Sigma_{\text{g}}^{\text{+}}$ to $\text{A}^{\text{2}} \Pi_{\text{u}}$ states~\cite{zhang19}. The onset of ionization and plasma defocusing may explain lower gain at high control intensity in Fig.~\ref{fig:wp}(c).

Neither the pump nor control pulse generate a strong rotational wave packet in the $\text{X}^{\text{2}} \Sigma_{\text{g}}^{\text{+}}$ state. If the $\text{X}^{\text{2}} \Sigma_{\text{g}}^{\text{+}}$ state is populated, the pulses may not generate a coherent wave packet because of the simultaneous interaction with the $\text{A}^{\text{2}} \Pi_{\text{u}}$ state. The bandwidth of the pulses covers more than three $\text{X}^{\text{2}} \Sigma_{\text{g}}^{\text{+}}$ to $\text{A}^{\text{2}} \Pi_{\text{u}}$ state vibronic transitions~\cite{gilmore92}. Figure~\ref{fig:levels}(a) shows that the equilibrium internuclear separation of the $\text{A}^{\text{2}} \Pi_{\text{u}}$ state is relatively large, so vibrational motion can temporarily trap population in the $\text{A}^{\text{2}} \Pi_{\text{u}}$ state. The pulse durations exceed the timescale of vibrational dynamics, so the interaction may scramble the phases in the $\text{X}^{\text{2}} \Sigma_{\text{g}}^{\text{+}}$ state as population moves between the two states and becomes temporarily trapped. This could quench rotational coherence in the $\text{X}^{\text{2}} \Sigma_{\text{g}}^{\text{+}}$ state as it forms. This is similar to an explanation of Pump--Probe--Control measurements~\cite{zhang19}, where the control pulse modifies the coherent emission between the $\text{B}^{\text{2}} \Sigma_{\text{u}}^{\text{+}}$ and $\text{X}^{\text{2}} \Sigma_{\text{g}}^{\text{+}}$ states.

\subsection{Pump--Probe--Control}

\begin{figure}[]
\includegraphics[trim={0cm 0cm 0cm 0cm},clip]{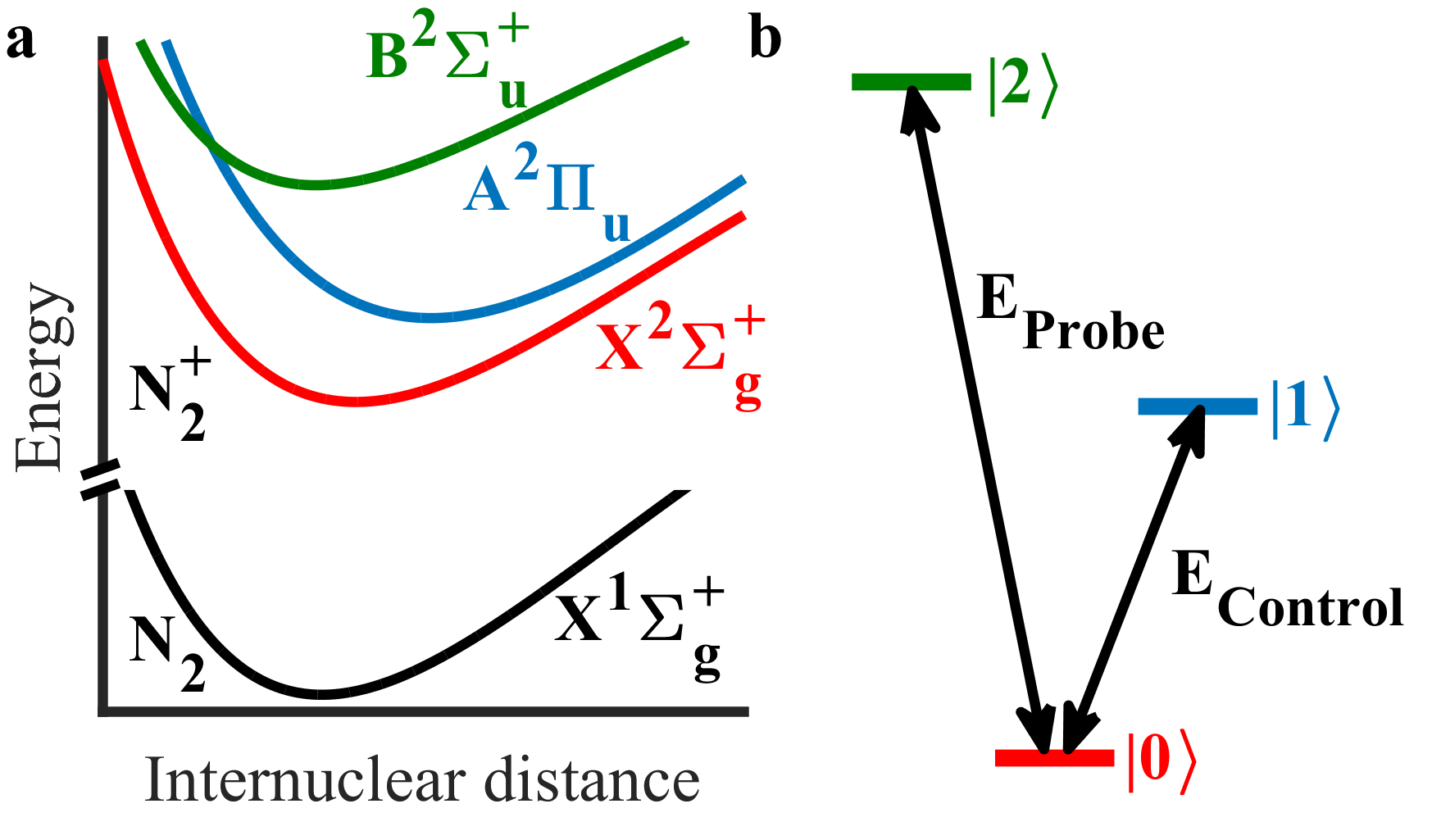}
\caption{\label{fig:levels} Energy diagrams. (a) Potential energy of the ground state of neutral nitrogen and the $\text{X}^{\text{2}} \Sigma_{\text{g}}^{\text{+}}$,  $\text{A}^{\text{2}} \Pi_{\text{u}}$, and $\text{B}^{\text{2}} \Sigma_{\text{u}}^{\text{+}}$ states of $\text{N}_{\text{2}}^{\text{+}}$. (b) Energy levels of a \textit{V}-system.}
\end{figure}

The probe pulse generates coherent emission at 391~nm [$\text{B}^{\text{2}} \Sigma_{\text{u}}^{\text{+}}$~$\left(\nu = 0\right)$~$\rightarrow$~$\text{X}^{\text{2}} \Sigma_{\text{g}}^{\text{+}}$~$\left(\nu = 0\right)$], and the control pulse reduces gain when it arrives during the emission in Fig.~\ref{fig:intro}. Therefore, the control pulse must modify the emission. The transition at 785~nm [$\text{A}^{\text{2}} \Pi_{\text{u}}$~$\left(\nu = 2\right)$~$\rightarrow$~$\text{X}^{\text{2}} \Sigma_{\text{g}}^{\text{+}}$~$\left(\nu = 0\right)$] forms a \textit{V}-system with the transition at 391~nm, as illustrated in Fig.~\ref{fig:levels}(b). Like the rotational coherence in the $\text{X}^{\text{2}} \Sigma_{\text{g}}^{\text{+}}$ state, an 800~nm pulse can modify the coherent emission between the $\text{X}^{\text{2}} \Sigma_{\text{g}}^{\text{+}}$ and $\text{B}^{\text{2}} \Sigma_{\text{u}}^{\text{+}}$ states by coupling the $\text{X}^{\text{2}} \Sigma_{\text{g}}^{\text{+}}$ and $\text{A}^{\text{2}} \Pi_{\text{u}}$ states because of the shared ground state. This case is slightly different because separate pulses create and modify the coherence, and the excited states cannot directly couple. 

\begin{figure}[]
\includegraphics[trim={0cm 0cm 0cm 0cm},clip]{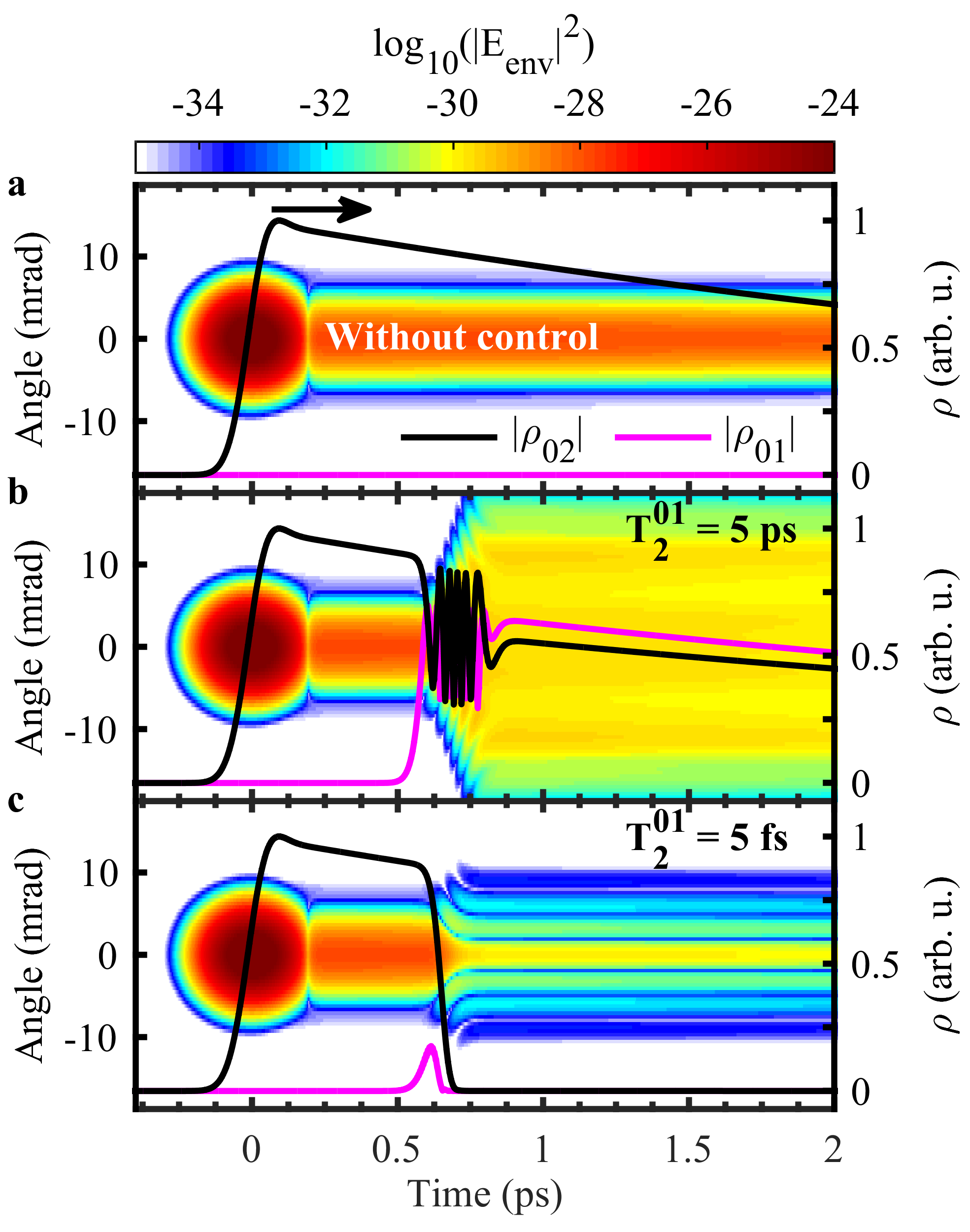}
\caption{
\label{fig:sim} 
Calculation results. (a) The spatial intensity profile of the emission on a logarithmic colour scale without the control pulse for reference. The superimposed lines show the magnitudes of the off-diagonal density matrix elements at the center of the beam as a function of time (right-hand axis). (b) The control pulse modifies the wavefront of the emission when it arrives at 0.7~ps. (c) Quenching dominates using $T_2^{01} = 5$~fs.}
\end{figure}

The control pulse can modify the emission in two distinct ways that we illustrate using a numerical three-level density matrix calculation. Figure~\ref{fig:sim}(a) shows the calculated spatial profile of the probe pulse as a function of time and divergence on a logarithmic colour scale without the control pulse. The superimposed lines show the magnitudes of the off-diagonal density matrix elements. The probe pulse couples $\ket{0}$ and $\ket{2}$ weakly, which generates coherence $\rho_{02}$ and emission trailing the pulse. The phenomenological coherence time ($T_2^{02} =$~5~ps) determines the linewidth and the duration of the emission.

The control pulse arrives at 0.7~ps and strongly couples $\ket{0}$ and $\ket{1}$ in Fig.~\ref{fig:sim}(b) and (c). In Fig.~\ref{fig:sim}(b), the coherence time of the transition near 800~nm is $T_2^{01} = 5$~ps. Rabi oscillations between $\ket{0}$ and $\ket{1}$ produce coherence $\rho_{01}$ that grows and oscillates, and they also modify $\rho_{02}$ because of the shared ground state. This induces a phase shift in $\rho_{02}$ and the resulting emission. The spatial profile of the control pulse makes the interaction non-uniform, so Rabi oscillations vary in the plane perpendicular to the propagation direction. The spatially dependent phase shift modifies the wave front and divergence of the emission like a spatial light modulator~\cite{li18}. Similarly, the Stark shift induced by a nonresonant control pulse was used to redirect extreme ultraviolet free induction decay~\cite{bengtsson17}. The phase shift due to the dynamic Stark effect was relatively small in this calculation, so that contribution is excluded.

The coherence time is $T_2^{01} = 5$~fs in Fig.~\ref{fig:sim}(c), which could represent the time limit for population transfer with the middle state before vibrational motion temporarily traps the population. The control pulse generates damped Rabi oscillations between $\ket{0}$ and $\ket{1}$ and short-lived coherence $\rho_{01}$ due to the fast coherence time. In the \textit{V}-system, $\rho_{02}$ decays with $\rho_{01}$ and this quenches the emission. 

In both cases, the control pulse also abruptly modifies the population of $\ket{0}$ during the emission, which generates a sharp feature in the emission and adds oscillations to the entire probe pulse spectrum that are highlighted in Supplementary Fig.~S3~\cite{SM}.

These results are consistent with Pump--Probe--Control experimental results. After the probe pulse measures gain, the spectrometer integrates over the emission without temporal information. These results show that the control pulse can quench or redirect the remaining emission depending on the coherence time of the $\text{X}^{\text{2}} \Sigma_{\text{g}}^{\text{+}}$ and $\text{A}^{\text{2}} \Pi_{\text{u}}$ states, which reduces the intensity at the on-axis detector. The control pulse influences the $\text{N}_{\text{2}}^{\text{+}}$ air laser pulse duration, spectrum, and direction.

\subsection{Summary}

In conclusion, we performed pump-probe experiments on the nitrogen molecular cation but labeled the excitation pulse ``control'' for clarity. Rotational excitation by the control pulse removed the initial rotational wave packet from the $\text{B}^{\text{2}} \Sigma_{\text{u}}^{\text{+}}$ state. We never observed a strong rotational wave packet in the $\text{X}^{\text{2}} \Sigma_{\text{g}}^{\text{+}}$ state. We proposed that this is due to coupling around 800~nm by the pump and control pulses that mixes the $\text{X}^{\text{2}} \Sigma_{\text{g}}^{\text{+}}$ and $\text{A}^{\text{2}} \Pi_{\text{u}}$ states. Our measurements led to a three-level \textit{V}-system model. It showed that coherent emission following the probe pulse is quenched or modified by the control pulse coupling the ground and middle states. Our measurement showed that emission is indeed interrupted by the control pulse.

Removing the initial rotational wave packet generates a smoother gain decay, but this has consequences for the emission. The emission also contains structure because of modulated gain during propagation and beating between transitions from different rotational states~\cite{zhong17,zhong18,miao18}, but the control pulse should also smooth the temporal profile of the emission. In addition, the ability to quench or coherently redirect the emission allows the control pulse to tune the characteristics of the air laser, which could be achieved from a standoff distance. The control pulse spatial profile, temporal shape, and polarization add more degrees of freedom to craft desired behaviour.

The major unresolved issue in $\text{N}_{\text{2}}^{\text{+}}$ air lasing is the detailed mechanism to build inversion. It seems likely to involve the $\text{A}^{\text{2}} \Pi_{\text{u}}$ state  as a population sink. An obvious next step is to change the pump wavelength away from $\text{X}^{\text{2}} \Sigma_{\text{g}}^{\text{+}}$ to $\text{A}^{\text{2}} \Pi_{\text{u}}$ state transitions so tunnel ionization establishes the initial populations in the ion. Then, only the control pulse will couple them, and it can manufacture the inversion using the $\text{A}^{\text{2}} \Pi_{\text{u}}$ state dynamics.

\begin{acknowledgements}
The authors are grateful for discussions with Michael Spanner, Misha Ivanov, Felipe Morales, Maria Richter, Pavel Polynkin, and Andrei Naumov. This research is supported by the U.S. Army Research Office through Award No. W911NF-14-1-0383.
\end{acknowledgements}

\section*{Appendix A: Experiment details}
The experimental setup adds the control pulse to a previous setup~\cite{britton18,britton19}. The input laser (800~nm) is split into three paths to form two Mach-Zehnder-type interferometers. Linear translation stages change the length of two paths to set the delay of the probe and control pulses relative to the pump pulse. Zero probe and control delays correspond to the arrival of the pump pulse in the focus. A 200~$\mu$m thick BBO crystal in the probe path frequency-doubles the probe pulse, and spectral filters isolate the second harmonic from the fundamental. A dichroic mirror recombines the probe pulse with the pump and control pulses. A half wave plate and polarizer in each path control the intensity and polarization of each pulse. 

The polarizations of the pump and probe are linear and parallel. A quarter wave plate in the control path changes the polarization of the control pulse to near circular ($\epsilon \approx 0.7$) after the polarizer. The three pulses are collinear. In this geometry, molecular alignment is different during the revival of pump- and control- induced rotational wave packets. The circularly polarized control pulse generates alignment along the propagation direction during the revival~\cite{smeenk13}, while the pump pulse generates alignment along the polarization direction during the revival. In both cases, the rotational wave packets modulate alignment in three dimensions. The probe pulse is sensitive to alignment along the polarization direction.

A concave mirror ($f/30$) focuses the pulses into the $\text{N}_{\text{2}}$ gas jet in vacuum. A 200~$\mu$m wide sonic nozzle and a pulsed valve with a backing pressure of {\raise.17ex\hbox{$\scriptstyle\mathtt{\sim}$}}6~bar of nitrogen gas generate the supersonic jet. Operating pressure in the chamber is {\raise.17ex\hbox{$\scriptstyle\mathtt{\sim}$}}$10^{-5}$~mbar. Three linear stages control the position of the nozzle relative to the laser in the vacuum chamber. The pump pulse ionizes $\text{N}_{\text{2}}$ and creates a plasma channel in the jet. A nearby conductive mesh measures ionization, which is maximum at the location of the focus. The nozzle is {\raise.17ex\hbox{$\scriptstyle\mathtt{\sim}$}}250~$\mu$m upstream from the focus, so the nozzle enclosure does not obscure the laser. The pump pulse creates high harmonics in the focus, which are collected on an XUV spectrometer in vacuum. The cut-off in the high harmonic spectrum provides the pump pulse intensity. The pump intensity determines the intensities of probe and control pulses using their relative energy, duration, and size.

The probe pulse is amplified after the pump pulse ionizes the jet. A mirror on a translation stage directs the laser out of the vacuum chamber. Absorption filters, interference filters, and dichroic mirrors isolate the probe pulse. A lens refocuses the probe onto a conventional UV/Vis fiber spectrometer (Ocean Optics Maya Pro 2000) that measures the spectrum. The amplification ratio is the intensity integrated over the peak in spectrum that roughly corresponds to the P-branch at 391~nm ($I_{out}$), which is divided by a reference value of the integrated intensity with no gain present ($I_{in}$). The natural logarithm of the amplification ratio is the gain--length product 

\begin{equation}
gL = log(\frac{I_{out}}{I_{in}}).
\end{equation}

\noindent
The modulations are isolated from decaying $gL$ by fitting and subtracting an exponential decay function. The modulations are multiplied by a broad sinusoidal function, so their amplitude smoothly decreases to zero at the start and end, and both ends are padded with zeros to improve the Fourier transforms. The Fourier transforms of Pump--Control--Probe measurements only include data from probe delays after the control delay.

\section*{Appendix B: Calculation details}
The calculation considers three states in a V configuration ($\ket{0}, \ket{1},$ and $\ket{2}$). The Hamiltonian for this system is $H = H_0 + V(t)$, where
\begin{equation}
\label{eq:H0}
H_0 = 
\begin{pmatrix}
	0 & 0 & 0 \\
	0 & \epsilon_1 & 0 \\
	0 & 0 & \epsilon_2 \\
\end{pmatrix},
\end{equation}

\noindent
and
\begin{equation}
\label{eq:V}
V(t) =
\begin{pmatrix}
	0 & -\mu_{01} E_C(t) & -\mu_{02} E_P(t) \\
	-\mu_{01} E_C(t) & 0 & 0 \\
	-\mu_{02} E_P(t) & 0 & 0 \\
\end{pmatrix}.
\end{equation}

\noindent
In the $\text{N}_{\text{2}}^{\text{+}}$ system, the transition dipole moments for the $\text{B}^{\text{2}} \Sigma_{\text{u}}^{\text{+}}$ to $\text{X}^{\text{2}} \Sigma_{\text{g}}^{\text{+}}$ transition and the $\text{A}^{\text{2}} \Pi_{\text{u}}$ to $\text{X}^{\text{2}} \Sigma_{\text{g}}^{\text{+}}$ transition are parallel and perpendicular, respectively. To compare with the calculated system, we effectively assume that the probe and control pulses have perpendicular polarization and that the molecules are constantly aligned with the probe polarization at a fixed internuclear distance.

The calculation uses atomic units (a.u.). The dipole moments are $\mu_{01} = 0.26$~a.u., $\mu_{02} = 0.74$~a.u., and the state energies are $\epsilon_1 = 1.54$~eV, and $\epsilon_2 = 3.17$~eV~\cite{gilmore92}. The initial electric field of the probe ($E_P$) and control ($E_C$) pulses is a transform-limited Gaussian pulse multiplied by a cosine function and set to zero past the first cosine zero-crossing. The cosine function provides a smooth decrease to zero, so the free induction decay is not seeded. The peak intensity of the probe pulse is $I_{P,0} = 10^8$~W/cm$^2$ and the control pulse is $I_{C,0} = 4\times10^{12}$~W/cm$^2$. Their full width at half maximum duration is 100~fs and the field is zero beyond $\pm$400~fs. The center wavelength of the probe (393~nm) and control (800~nm) pulses are detuned by a couple of nanometers from each transition. The time grid spans from -0.5~ps to 15~ps using a time step of 20~as.

The Liouville-von Neumann equation $i \hbar \dot{\rho} = \left[H, \rho\right]$ determines the evolution of the density matrix ($\rho$) in time. The initial populations are $\rho_{00,0} = 0.2$, $\rho_{11,0} = 0.1$, and $\rho_{22,0} = 0.7$, and population decays to the ground state on the timescale of $T_1 = 10$~ns. The results are qualitatively similar for a free induction decay due to absorption when the system is not inverted. The off-diagonal density matrix elements are the coherence between states, so they are initially zero. The coherence decay time for the $\ket{0}$ and $\ket{2}$ states is $T_2^{02} = 5~\mathrm{ps}$, which is similar to the 391~nm emission duration~\cite{yao13,li14,liu15,zhong17,zhong18}. We change the coherence decay time for the $\ket{0}$ and $\ket{1}$ states ($T_2^{01}$) to observe different behaviours. The Crank-Nicolson method in time provides a numerical solution to the equations of motion for the density matrix elements~\cite{crank47}. 

The material polarization $P_{\mu} = N\mu_{02}(\rho_{02} + \rho_{20})$, where $N$ is the number density, is a source term for the probe pulse propagation using the one-direction decomposition of the wave equation in the reference frame of the probe pulse 

\begin{equation}
\frac{\partial{E_P(t,z)}}{\partial{z}} = -(\frac{1}{c} - \frac{1}{v_g}) \frac{\partial{E_P(t,z)}}{\partial{t}} - \frac{2\pi}{c} \frac{\partial{P_{\mu}(t,z)}}{\partial{t}},
\end{equation}

\noindent
where $v_g$ is the group velocity. Similarly, $\rho_{01}$ is included in the control pulse propagation. The propagation step of 1~$\mu$m in a density of $10^{17}$~cm$^{-3}$ creates a small free induction decay. Propagation includes first and second order changes to the electric field and is suitable for many propagation steps, but the calculation uses a single step to accommodate the computation time required for many near-field positions.

The probe and control pulses are collinear and have beam waists of 50 and 75~$\mu$m, respectively. The near-field radial axis has 500 positions and extends to 499~$\mu$m. The particular radial positions enable an efficient implementation of the Hankel transform, which converts the electric field from the near field radial position to the far field divergence at each time step~\cite{guizar04}. The emission is calculated for the first 200 positions, while the remaining are padded with zeros. The closest on-axis radial position is 0.8~$\mu$m. The far-field divergence axis assumes the transition frequency corresponding to 391~nm but depends on frequency in general. This is a reasonable assumption for the free-induction decay. The Hilbert transform provides the electric field envelope, which improves the appearance of figures. 

The calculation showed that population transfer between $\ket{0}$ and $\ket{1}$ modifies the coherence between $\ket{0}$ and $\ket{2}$ independent of $T_2^{01}$, but $T_2^{01}$ determines whether the emission is coherently redirected or quenched. When coherent population transfer dominates, the emission contains a phase shift in time that corresponds to a different lineshape in frequency~\cite{ott13}. Other effects can also redirect and quench the emission, like the Stark shift of the levels by the control pulse~\cite{bengtsson17}. If the control pulse was detuned from resonances, Rabi flopping and Stark shifts by the control pulse would both be important contributions. 

\bibliographystyle{apsrev4-1}

%


\onecolumngrid
 \vspace{1.0cm}
 \clearpage

  \setcounter{section}{0}%
  \setcounter{subsection}{0}%

\onecolumngrid
\begin{center}
  \textbf{\large {Supplementary Materials for}\\
Control of $\text{N}_{\text{2}}^{\text{+}}$ Air Lasing}\\[.2cm]
 Mathew Britton,$^{1,*}$ Marianna Lytova,$^{1}$ Dong Hyuk Ko,$^{1}$ Abdulaziz Alqasem,$^{1,2}$ Peng\\
  Peng,$^{1}$ D. M. Villeneuve,$^{1,3}$ Chunmei Zhang,$^{1,\dagger}$ Ladan Arissian,$^{1,3,4}$ and P. B. Corkum$^{1,\ddag}$\\[.1cm]
  {\itshape ${}^1$Department of Physics, University of Ottawa, Ottawa, K1N 6N5, Canada\\
  ${}^2$Department of Physics, King Saud University, Riyadh, 11451, Saudi Arabia\\
  ${}^3$National Research Council of Canada, Ottawa, K1A 0R6, Canada\\
  ${}^4$Center for High Technology Materials, Albuquerque, NM, 87106, USA\\}
\end{center}
 \vspace{1.0cm}
\twocolumngrid

\setcounter{equation}{0}
\setcounter{figure}{0}
\setcounter{table}{0}
\setcounter{page}{1}

\renewcommand{\thefigure}{S\arabic{figure}}

\section{Rotational wave packets}

\begin{figure}[h]
\centering
\includegraphics[trim={0cm 0cm 0cm 0cm},clip]{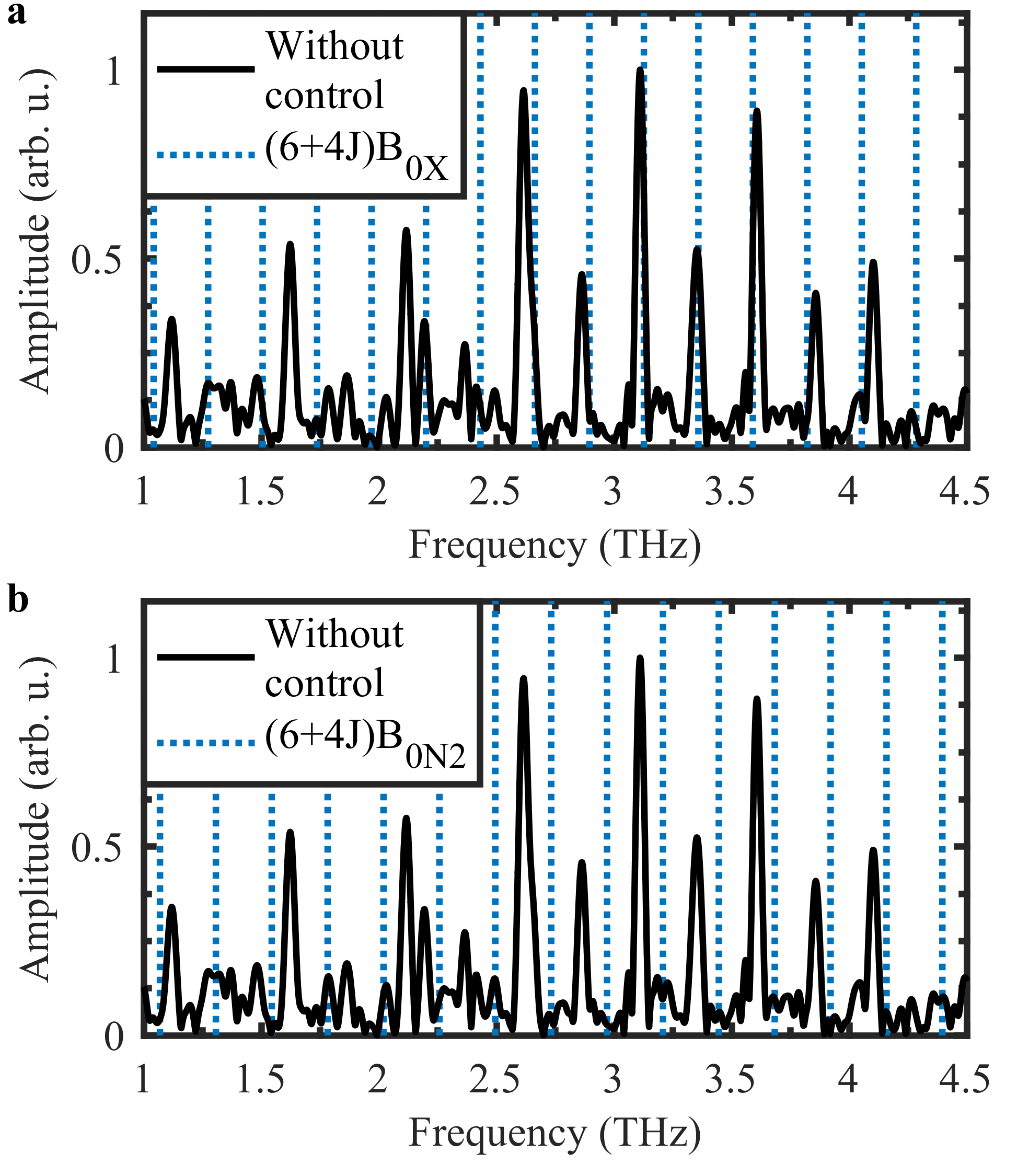}
\caption{\label{fig:fft} Amplitude of the Fourier transform of the modulations as a function of frequency without a control pulse. The vertical dashed lines indicate the rotational frequencies for the ground state of (a) the ion and (b) the neutral molecule. $I_{pump} = 2.5\times10^{14}~\mathrm{W}/\mathrm{cm}^2$. }
\end{figure}

The rotational wave packet in the $\text{B}^{\text{2}} \Sigma_{\text{u}}^{\text{+}}$ state is the dominant contribution to the modulations superimposed on the gain decay, which is visible by comparing the Fourier transform of the modulations with the rotational frequencies of each state. Supplementary Fig.~\ref{fig:fft} shows this comparison for the two states with relatively poor agreement: the ground state of the ion and the neutral molecule.

\section{Control-induced gain}

\begin{figure}[h]
\centering
\includegraphics[trim={0cm 0cm 0cm 0cm},clip]{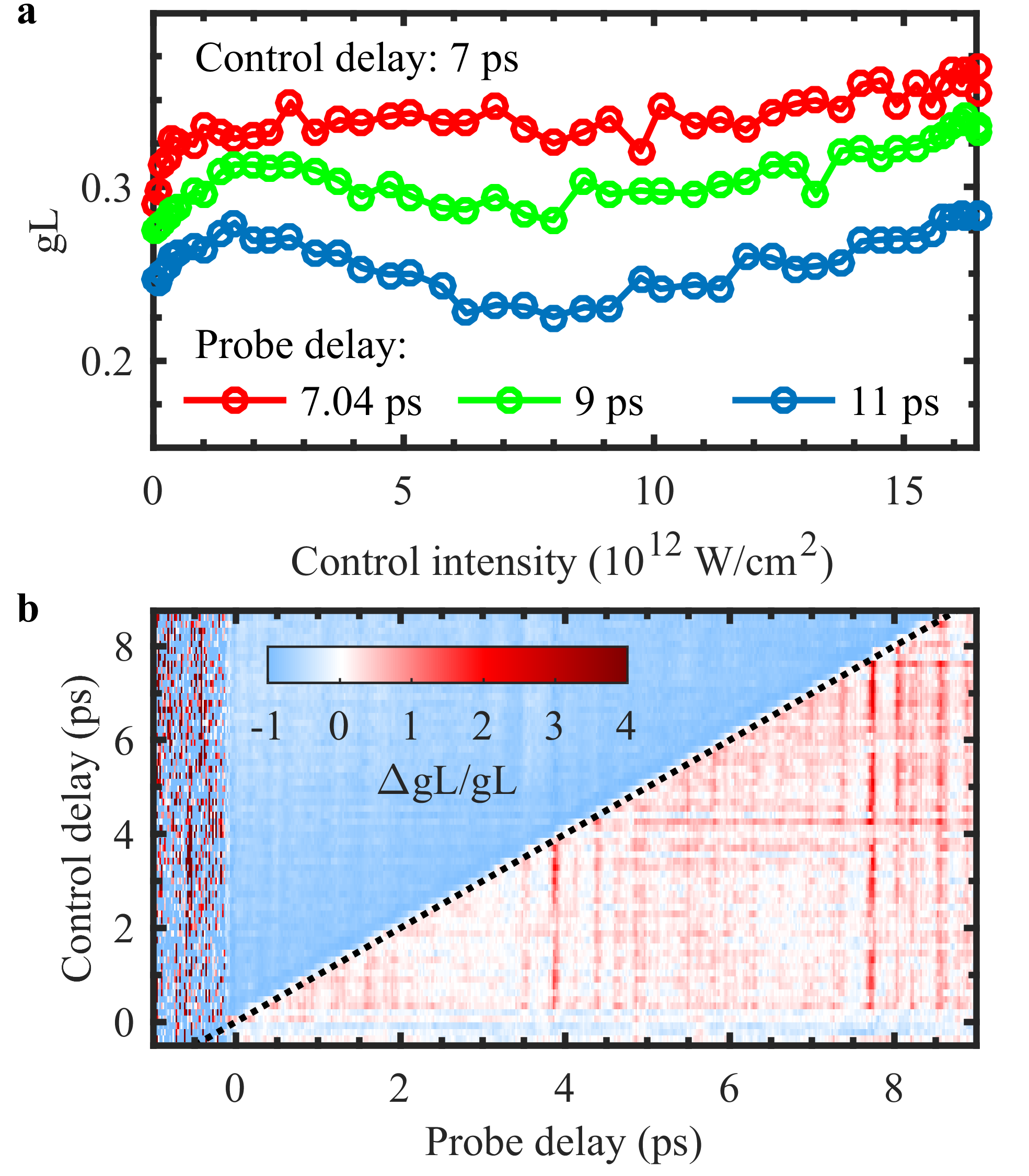}
\caption{\label{fig:gainchange}(a) Gain at three probe delays as a function of control intensity after the control pulse interaction. Gain at the lowest control intensity is approximately equal to gain with no control pulse. $I_{pump} = 1.9\times10^{14}~\mathrm{W}/\mathrm{cm}^2$. (b) The relative difference between gain with the control pulse and without control as a function of control and probe delay. $I_{pump} = 2.5\times10^{14}~\mathrm{W}/\mathrm{cm}^2$, $I_{control} = 10^{13}~\mathrm{W}/\mathrm{cm}^2$.}
\end{figure}

The control pulse exchanges population between the $\text{X}^{\text{2}} \Sigma_{\text{g}}^{\text{+}}$ and $\text{A}^{\text{2}} \Pi_{\text{u}}$ states, which may destroy rotational coherence or modify the emitting system. This is accompanied by a control-induced change of gain due to the population transfer involving the ground state. Supplementary Fig.~\ref{fig:gainchange}(a) shows gain at a fixed control delay and three later probe delays (\textit{i.e.} Pump--Control--Probe). The control-induced change is apparent as the intensity increases. There is an upward trend as a function of intensity, in addition to a dip at higher intensity. The dip is likely from modified rotational wave packets, as it appears at intensities where the control pulse begins to modify modulations. Overall, gain is slightly higher after interacting with the control pulse, but this also depends on control delay.

\begin{figure}[]
\centering
\includegraphics[trim={0cm 0cm 0cm 0cm},clip]{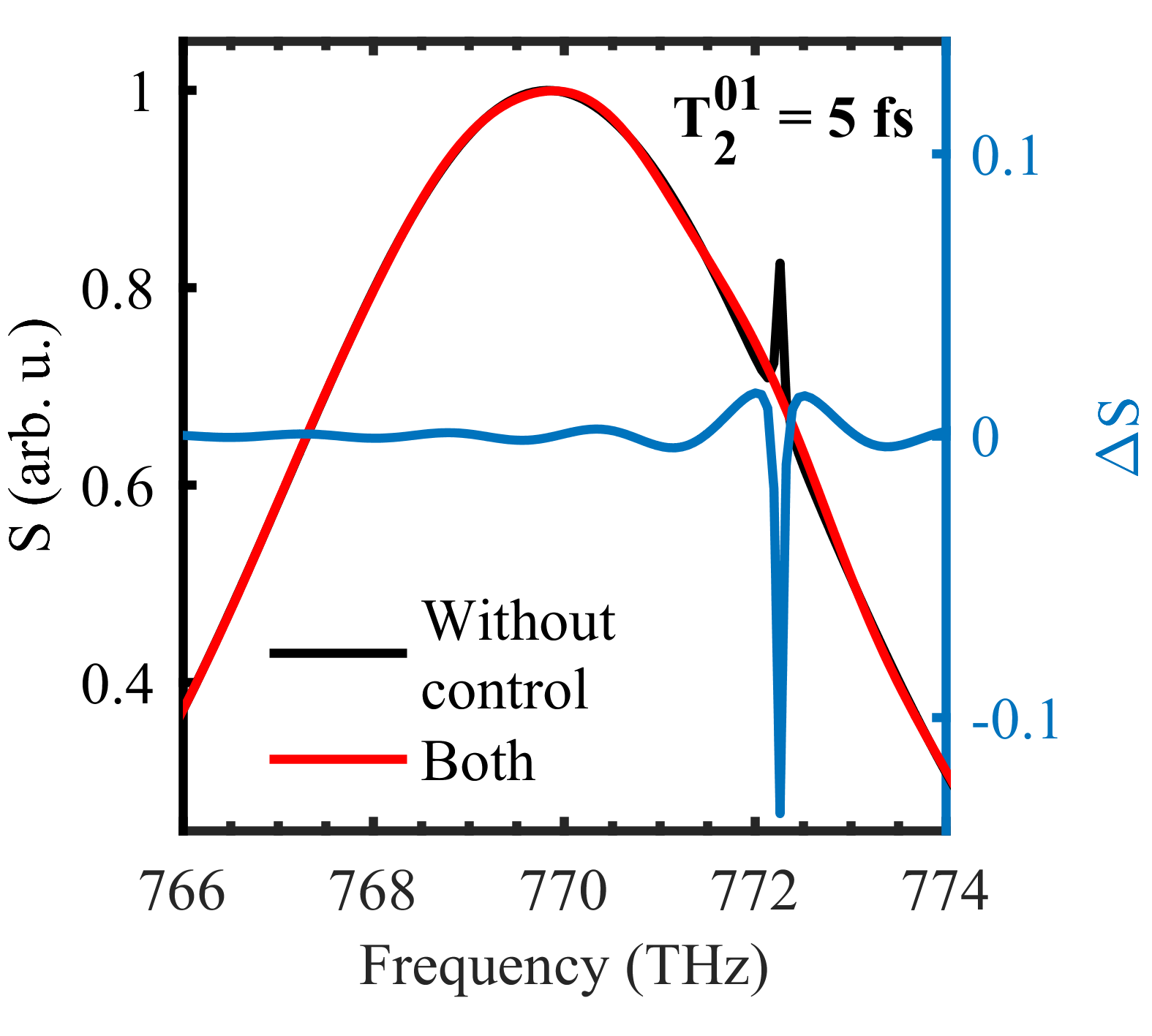}
\caption{\label{fig:spectraloscillations} The calculated spectrum of the probe pulse with and without the control pulse, and the difference on a separate axis ($\Delta S$).}
\end{figure}

Supplementary Fig.~\ref{fig:gainchange}(b) shows the control-induced change to gain

\begin{equation*}
\Delta\text{gL}/\text{gL}= \frac{\text{gL}_{\text{Control}} - \text{gL}}{\text{gL}}
\end{equation*}

\noindent
as a function of probe and control delay, which is the relative difference between gain with the control pulse ($\text{gL}_{\text{Control}}$) and without control ($\text{gL}$). The diagonal line is the separation between Pump--Probe--Control (top) and Pump--Control--Probe (bottom) measurements. The negative top region is modified emission. The Pump--Control--Probe region is mostly positive with additional structure, so gain is higher on average in the presence of the control pulse at this intensity.

\section{\textit{V}-system spectral oscillations}

A simple three-level model of a \textit{V}-system shows that exchanging population on one transition can modify the emission on the other transition. The emission is quenched or coherently modified depending on the coherence time during the population exchange. Regardless of the coherence time, gain is abruptly modified during the emission. This generates a sharp feature in the emission that corresponds to broader spectral bandwidth. As a result, the amplified spectrum contains oscillations that are highlighted in Supplementary Fig.~\ref{fig:spectraloscillations}. The oscillations can be described as spectral interference between the original emission and the control-induced change to the emission.

\end{document}